\documentclass[lettersize,journal]{IEEEtran}
\pdfoutput=1
\usepackage{amsmath,amsfonts}
\usepackage{algorithmic}
\usepackage{array}
\usepackage[caption=false,font=normalsize,labelfont=sf,textfont=sf]{subfig}
\usepackage{textcomp}
\usepackage{stfloats}
\usepackage{url}
\usepackage{verbatim}
\usepackage{graphicx}
\usepackage[capitalize,noabbrev,nameinlink]{cleveref}
\usepackage{amsmath,amsfonts,amssymb,amsthm}
\usepackage{parskip}
\usepackage{graphicx,nicefrac}
\usepackage{float}
\usepackage{enumitem}
\usepackage{xcolor}
\usepackage{cancel}
\usepackage{algorithm}
\usepackage{algorithmic}
\usepackage{tabularx}
\usepackage{multirow}
\usepackage[font=small]{caption}
\usepackage{cleveref}

\usepackage[backend=biber,style=numeric-comp,sorting=none,maxbibnames=99]{biblatex}
\usepackage{balance}
\usepackage{dirtytalk}
\usepackage{afterpage}

\usepackage{times}

\newcommand{\R}{\mathbb{R}}

\newcommand{\dyn}{f}

\newcommand{\state}{x}

\newcommand{\control}{u}

\title{\LARGE \bf Decomposing Control Lyapunov Functions\\ for  Efficient Reinforcement Learning }

\author{%
  \IEEEauthorblockN{%
    Antonio L\'opez\textsuperscript{1}
    and David Fridovich-Keil\textsuperscript{1}
  }%
}
\begin{document}

\maketitle
\begingroup\renewcommand\thefootnote{1}
\footnotetext{Department of Aerospace Engineering and Engineering Mechanics, University of Texas at Austin. \{\texttt{alopezz, dfk}\}\texttt{@utexas.edu}.\\
\hspace*{4mm}This work was supported by the Fulbright U.S. Student Program, which is sponsored by the U.S. Department of State and Fulbright Comexus. Its contents are solely the responsibility of the authors and do not necessarily represent the official views of the Fulbright Program.\\
\hspace*{4mm}This research was also sponsored by the Army Research Laboratory and was accomplished under Cooperative Agreement Number W911NF-23-2-0011.}
\endgroup
\thispagestyle{empty}
\pagestyle{empty}

%\textbf {\emph{Abstract{---}}
\hspace*{4mm}
\begin{abstract}
Recent methods using Reinforcement Learning (RL) have proven to be successful for training intelligent agents in unknown environments. However, RL has not been applied widely in real-world robotics scenarios. This is because current state-of-the-art RL methods require large amounts of data to learn a specific task, leading to unreasonable costs when deploying the agent to collect data in real-world applications. In this paper, we build from existing work that reshapes the reward function in RL by introducing a Control Lyapunov Function (CLF), which is demonstrated to reduce the sample complexity. Still, this formulation requires knowing a CLF of the system, 
but due to the lack of a general method, it is often a challenge to identify a suitable CLF. Existing work can compute low-dimensional CLFs via a Hamilton-Jacobi reachability procedure. However, this class of methods becomes intractable on high-dimensional systems, a problem that we address by using a system decomposition technique to compute what we call Decomposed Control Lyapunov Functions (DCLFs). We use the computed DCLF for reward shaping, which we show improves RL performance. Through multiple examples, we demonstrate the effectiveness of this approach, where our method finds a policy to successfully land a quadcopter in less than half the amount of real-world data required by the state-of-the-art Soft-Actor Critic algorithm.
\end{abstract}

\vspace{-6pt}
\section{Introduction}
\hspace*{4mm}Learning control policies for autonomous robots in complicated environments has many applications, such as geographic mapping with drones or exploration on dangerous terrain with quadrupeds [1]. Finding control policies for these robots is challenging, as these systems have complex nonlinear dynamics, and there is almost never complete information to model the environment where the robot is being deployed [2]. In these situations, data-driven approaches are used to account for uncertainties in the model. 

\vspace{-6pt}
\hspace*{4mm}Reinforcement Learning (RL) has proven to be an effective data-driven method for devising policies in unknown environments. However, the effectiveness of RL algorithms heavily depends on the design of the reward function, tuning of hyper-parameters, and the data available [3]. Although RL algorithms can identify near-optimal policies, they often require extensive data sets for policy training [32], which can be impractical in real-world setups as the cost to collect data becomes high. 

\begin{figure}[!htbp]
    \begin{center}
         \includegraphics[width=\linewidth]{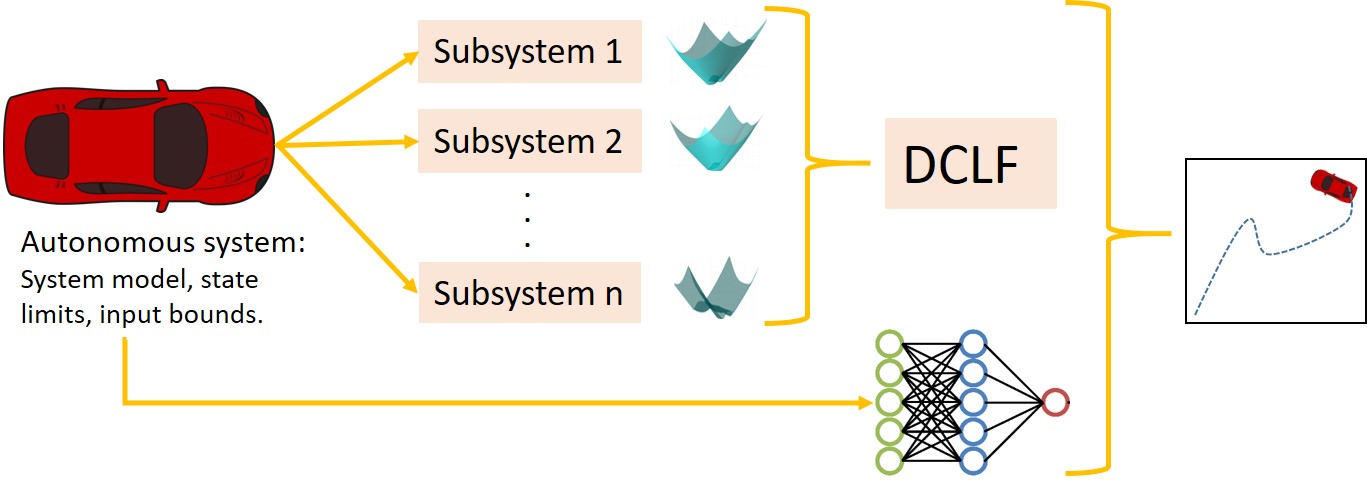}
    \end{center}
\caption{Schematic of our framework. We take a dynamical system model of an autonomous robot and decompose it into several subsystems. We compute a Control Lyapunov-Value Function (CLVF) for each subsystem and take the sum of these CLVFs as our Decomposed Control Lyapunov Function, which we show can be incorporated as reward shaping to accelerate policy learning in a variety of low- and high-dimensional tasks.}
\label{Fig1}
\end{figure}

\vspace{-6pt}
\hspace*{4mm} A recent approach [31] addresses the high-sample complexity of RL algorithms by introducing a Control Lyapunov Function (CLF) in the reward function.  
This incentivizes policies that improve the value of the CLF along the trajectory. Since a CLF captures the stability properties of a system, this reformulation will incentivize policies that stabilize the system, helping in the exploration process of RL by requiring less data to find a stabilizing controller. The work [31] assumes that at least an approximate CLF for the system is already known. Unfortunately, there is no general procedure to find a CLF.\\ 
\hspace*{4mm}Existing methods can compute a type of CLF, called a Control Lyapunov-Value Function (CLVF), via Hamilton-Jacobi Reachability (HJR) analysis. Nonetheless, as the method uses HJR, it is prey to the ``curse of dimensionality," and the computation of the CLVF becomes intractable for systems with more than five dimensions [28], which is the case for most autonomous systems deployed in the real world.\\   
\hspace*{4mm}In this paper, we extend the method from [4] to find such functions for higher-dimensional systems. Using a system decomposition technique, we derive a Decomposed Control Lyapunov Function (DCLF) for higher-dimensional systems that can be incorporated into any standard RL algorithm for reward shaping.
 
\vspace{-6pt}
\hspace*{4mm} Fig. 1 shows a schematic of our approach, where we decompose the dynamical system of an autonomous robot into several ``subsystems," each one of 5 dimensions or less, where using HJR is feasible. Then, we obtain a CLVF for each subsystem, solving the corresponding HJR problem.
Next, these CLVFs are used to derive a DCLF on the full system. Finally, we use the DCLF for reward shaping in RL to obtain a policy on a complete system representation of a robot's dynamics. The main contributions of this work are as follows: 
\begin{itemize}
     \item We extend prior work on HJR to compute a DCLF for high-dimensional state spaces on a particular class of dynamical systems, providing a framework that can reshape the reward function in any standard RL algorithm. 
     
    \item We present several experiments to demonstrate the effectiveness of the DCLF in alleviating the sample complexity in RL. 

\end{itemize}
\section{Related work}
\hspace*{4mm}As traditional RL algorithms suffer from high sample complexity, recent approaches [5]-[8] focus on reducing the dependency of RL on large data sets in the training process.
Although these methods excel on their own without requiring any knowledge of the dynamical model, it is known that model-free methods can benefit when information about the dynamics is given [38]. 

%\vspace{-6pt}
%\hspace*{4mm}Although the previous methods mentioned excel on their own without requiring any knowledge of the model dynamics, it is known that model-free methods can benefit when information about the dynamics is given [8]. For example, Franke et al. [6] use a dynamics model to estimate the long-term value of the Q-function in traditional Q-learning, which reduces the sample complexity of learning. Even though this method was shown to learn policies faster, the asymptotic performance of the system is not discussed, which is a common disadvantage when introducing model-based approaches [18].  

\vspace{-6pt}
\hspace*{4mm} 
Existing literature [9]-[12] has incorporated information about the model dynamics by joining optimal control methods with reinforcement learning. The work from Nagami and Schwager [13] derives a policy on a simplified drone model by solving the corresponding Hamilton-Jacobi-Bellman (HJB) equation on a minimum-time optimal control problem. The authors use the policy to initialize the training of a policy-gradient RL algorithm, which uses a complete representation of the drone dynamics. Their approach is demonstrated to outperform model-
free RL methods, but is specific to systems when the dynamics can be modeled by a low-dimensional system only, where solving the HJB equation is feasible.
Similar to our approach in incorporating HJR analysis, Fisac et al. [14] combine HJR analysis with RL to make RL possible for safety purposes. This approach is shown to work on high-dimensional systems and opens a new avenue to join control theory and RL, but it can still benefit from the training efficiency of our approach. 

\vspace{-6pt}
\hspace*{4mm} An alternative way to incorporate domain knowledge into RL is reward shaping. This technique modifies the reward function to guide the RL algorithm faster toward more promising solutions [19]. One of the most well-known reward shaping methods is potential-based reward shaping [20]. This method, along with other variants [21-24], incorporates domain knowledge in the form of a potential-based function and assumes that the gradients of the potential function are aligned with directions of policy improvement. However, potential functions are typically designed based upon heuristics [20] and do not always lead to improvements in data efficiency [19]. In contrast, the work from [31] effectively incorporates domain knowledge as a CLF and shows improvement in the sample complexity of RL. As mentioned in the previous section, this method will be the basis of our approach.

\section{Preliminaries}

\hspace*{4mm}In this paper we consider autonomous systems with state $x \in \mathcal{X} \subset \R^n$ and control input $u \in \mathcal{U} \subset \R^m$, whose dynamics evolve according to the following equation: 
\begin{equation}\label{dyn}
    \dot{x}(t)=f(x(t),u(t)).
\end{equation}
We also assume the dynamics $f$ to be Lipschitz continuous in $(x,u)$ and consider time $t \in [t_0,0]$ with $t_0<0$, per the standard assumptions in HJR analysis [28]. Given these assumptions, initial state $x(t_0)$ and control $u(\cdot)$, the existence of a unique solution to the differential equation in (1) is guaranteed [35], which we define as the state trajectory $\beta(t)$.

\vspace{-6pt}
\subsection{Lyapunov Stability}
\label{Section:Lyap_stab}

\hspace*{4mm}For system (1), a CLF is a function $P:\mathcal{X} \times \mathcal{U}  \rightarrow \mathbb{R}$ that satisfies the following conditions in a neighborhood $\mathcal{O}$ around an equilibrium point $\bar{\state}$: 
\begin{enumerate}
   
    \item $P$ is positive definite, and continuously differentiable.
    \item For all $x$ in the neighborhood $\mathcal{O}$, there exists an input ${\control \in \mathcal{U}} \text{ such that } \dot{P}(\state) = \nabla{P} \cdot f(x,u) < 0.$ 
\end{enumerate}

\hspace*{4mm}The existence of a CLF implies that the system is asymptotically stabilizable to an equilibrium point [35]. If in addition, $\dot{V}(\state) \leq -\alpha V$ for some constant $\alpha > 0$ and all $x$, then exponential stabilizability is also guaranteed [35].

\subsection{Reinforcement learning} 
\hspace*{4mm}Consider the discrete-time version of (1) as: 
\begin{equation}
    x_{k+1}=f(x_k,u_k), 
\end{equation}
and define a positive definite cost function $\ell:\mathcal{X} \times \mathcal{U} \rightarrow \mathbb{R}$. Let $\Pi$ be the set of admissible policies $\pi:\mathcal{X} \rightarrow \mathcal{U}$, $x_0$ an initial state and $\lambda \in [0,1]$ a discount factor, then the associated value function is defined as: 
 \begin{equation}
     V_\lambda^\pi(x_0)= \sum_{k=0}^{\infty} \lambda^k\ell(x_k,\pi(x_k)).
 \end{equation}

 \vspace{-6pt}
\hspace*{4mm}RL algorithms seek to minimize the value function over the set of admissible policies. Unfortunately, it is unrealistic to search all possible policies. Instead, RL algorithms generally do this by randomly sampling trajectories of the system and updating the policy to decrease the long-run cost. The discount factor plays an important role in these algorithms since it represents how far into the future the algorithm plans. Larger $\lambda$'s effectively incorporate the cost in the distant future and, therefore, tend to require more training data.

\vspace{-6pt}
 \hspace*{4mm}Westenbroek et al. [31] proposed a modification of the value function, adding a known or approximate Lyapunov function of the system. Let a positive definite function ${P: \R^n\rightarrow \R}$ be a Lyapunov function of the dynamical system described in equation (1), then the value function is redefined as: 
 \begin{equation}\
         V_\gamma^\pi(x_0)= \sum_{k=0}^{\infty} \gamma^k[\Delta P(x_k)
         +\ell\big(x_k,\pi(x_k)\big)],
 \end{equation}
 where $\Delta P(x_k):=P\big(f(x_k,\pi(x_k))\big)-P(x_k)$.
 
\hspace*{4mm}Note that this modification will motivate policies that decrease the value of $P$ on the next state when applying an action at $x_k$, as this would decrease the overall cost. 

 \vspace{-6pt}
\subsection{Hamilton-Jacobi Reachability}
\hspace*{4mm}HJR analysis is a safety formulation that can be used for computing a Backward Reachable Set (BRS) given a specific target $\mathcal{T} \subset \mathbb{R}^{n}$ and time $t$. The BRS is the set of states from which the system can avoid the target set for all $t \in [t_0,0]$ by choosing an appropriate input from the set of admissible controls. To compute the BRS, we first define a Lipschitz continuous cost function $g:\R^n \rightarrow \R$ that satisfies $g(x)\geq 0$ only when $x \in \mathcal{T}$. Such a function can be the signed distance to the target set. The value function can then be defined as: 
\begin{equation}
    V(x,t)=\min_{u \in \mathcal{U}} \max_{t \in [t_0,0]} g(\beta(t)).
\end{equation}
\hspace*{4mm}This value function captures the maximum cost achieved over time, given that the optimal control was applied at all times. The zero super-level set of this value function is the set of states from which the system cannot avoid collision, i.e., cannot reach the target, whereas the BRS can be recovered as the zero sub-level set of $V$. The value function can be computed via dynamic programming by solving the following Hamilton-Jacobi-Isaacs (HJI) variational inequality [29]: 
\begin{equation}
    \min\biggl\{g(x) - V(x,t),\frac{\partial V}{\partial t}+\min_{u \in \mathcal{U}} \frac{\partial V}{\partial x} \cdot \dyn(x,\control)\biggl\} = 0, 
\end{equation}
with terminal value $V(x,0)=g(x)$.

Solving this differential variational inequality is difficult due to the ``curse of dimensionality" presented when dealing with high-dimensional state spaces. Still, numerical tools [30] have been developed to compute solutions for systems with state spaces of 5 dimensions or less. Dealing with higher dimensions in a particular variant of (6) is one of the contributions of this paper. 

\subsection{Control Lyapunov-Value Functions (CLVFs)}
\hspace*{4mm}As first introduced in [4], a CLVF for the system in (1) can be computed via HJR analysis. A CLVF has properties similar to a CLF and can stabilize the system to an equilibrium point or the Smallest Control Invariant Set (SCIS) if the system does not admit an equilibrium point. The stability problem is posed as a safety problem where the system seeks to avoid all regions outside the SCIS. First, the value function can be redefined from (5) as: 
\begin{equation}
    V_\gamma(x,t)=\min_{u \in \mathcal{U}} \max_{t \in [t_0,0]} e^{\gamma(t-t_0)}g(\beta(t)),
\end{equation}
where $\gamma$ represents the desired decay rate.
Then, the CLVF is defined as follows, assuming that the limit exists:
\begin{equation}
    V_\gamma^\infty=\lim_{t_0 \rightarrow -\infty} V_\gamma (x,t_0).
\end{equation}
This limit can be computed by solving the following HJI variational inequality [4]: 
\begin{equation}
\label{eqn:clvfvi}
    \begin{aligned}
        \max\biggl\{& g(x) - V_\gamma^\infty(x),\\
        &\min_{u \in \mathcal{U}} \frac{\partial V_\gamma^\infty}{\partial x} \cdot \dyn(x,\control) + \gamma V_\gamma^\infty(x)\biggr\} = 0.
    \end{aligned}
\end{equation}
\hspace*{4mm}Instead of recovering a BRS like in HJR analysis, the solution from the previous equation is a CLVF that can be used as a CLF for stabilizing the system.

\section{Decomposed Control Lyapunov Value Functions }

\hspace*{4mm}This work considers a class of nonlinear autonomous systems following equation (1). We aim to drive the system to an equilibrium point, which can be assumed to be the origin, since we can do a state transformation that maps the equilibrium point to the origin [35]. Completing seemingly simple tasks may require millions of steps when employing conventional RL algorithms. Moreover, including high-dimensional state spaces requires even more steps for data collection [32].
To reduce the required data, we build upon the method of [31] in which a CLF is incorporated in the reward function of the RL algorithm. However, we are dealing with high-dimensional spaces where there is no general method to compute a CLF for the system. In this section, we present our method to compute CLFs for high-dimensional systems and provide examples to show their effectiveness when incorporated into RL algorithms.  

\vspace{-6pt}
\subsection{System Decomposition}
\hspace*{4mm}Since HJR is computationally intractable for systems of more than five dimensions, a recent approach [33] decomposes a dynamical system into two or more lower-dimensional subsystems where HJR is feasible, that is, on systems with five dimensions or less. This decomposition is possible only when the subsystems can be coupled by some states or the controls. As an example, consider a system with dynamics given by:
\begin{equation}
    \dot{x}=f(x,u)=f(x_1,x_2,x_3,u) 
\end{equation}
where $x_1 \in \mathbb{R}^{n_1}, x_2 \in \mathbb{R}^{n_2}, x_3 \in \mathbb{R}^{n_3},  x \in \mathbb{R}^{n_1+n_2+n_3}$ and $u \in \mathcal{U}$. We say the system can be decoupled if we can partition the state $x$ into two groups: 
\begin{align}
    \begin{split}
        z_1&=(x_1,x_3)\\
        z_2&=(x_2,x_3)
    \end{split}
\end{align}
 and represent their dynamics into self-contained subsystems as in (12). 
 
\;\;\;\;\;\;\;\;\;\;\;\;\;\;\;Subsystem S1 \;\;\;\;\;\;\;\;\;\;\;\;\;\;\;\;Subsystem S2
\vspace{-6pt}
\begin{equation}
  \begin{split}
    \dot{z_1}&=g_1(x_1,x_3,u)\\
  \end{split}
  \;\;\;\;\;\;\;\;\;\;\;    
  \begin{split}
    \dot{z_2}&=g_2(x_2,x_3,u)\\
  \end{split}
\end{equation}

\vspace{-6pt}
\hspace*{4mm}Analogously, this partition can be done in more than two groups, and the theory still applies [33]. Each subsystem's dynamics are well-defined and do not depend on variables that are isolated within another subsystem; hence, they are self-contained. The subsystems share the states in $x_3$ (which may be an empty set, i.e., no overlapping states) and may or may not share the same control $u$. Later, we will show how this approach can work with and without shared controls on the subsystems.  

\vspace{-6pt}
\hspace*{4mm}We can solve a standard reachability problem to obtain the BRS for each subsystem and then take the back projection of each BRS onto the full-dimensional state. Finally, we can compute the full BRS by taking the intersection of the back-projected BRSs. 
Using a similar approach, we will show how to use system decomposition to obtain a type of CLF for high-dimensional systems instead of recovering reachable sets, as in [33], [36].

\subsection{Decomposed Control Lyapunov Functions}
\hspace*{4mm}Our method uses system decomposition to compute  Decomposed Control Lyapunov Functions (DCLF). First, we take the system dynamics described in equation (1) and decompose the system to obtain the form described in (12). Then, instead of recovering the BRS of the subsystems, we compute the CLVF for each subsystem and take the sum of these CLVFs as our DCLF. Formally, we define a DCLF as follows:

\vspace{-2pt}
\hspace*{4mm}\textbf{Definition 1.} Consider a system that can be written in the form as in (11). Suppose we have two CLFs satisfying the conditions described in \Cref{Section:Lyap_stab}, $V_1:\R^{n_1+n_3} \times \mathcal{U} \rightarrow \R$ and $ V_2: \R^{n_2+n_3} \times \mathcal{U} \rightarrow \R$ for the subsystems S1 and S2, respectively, as in (12). Then the function ${W(x_1,x_2,x_3,u)=V_1+V_2}$ is a DCLF. 

\vspace{-2pt}
\hspace*{4mm}\textbf{Proposition 1.} The function $W$ also satisfies the conditions from \Cref{Section:Lyap_stab}.
\vspace{-12pt}
\begin{proof} First note that the definition of $W$ does not restrict $V_1$ and $V_2$ to be in the same dimensional space, as both functions output a real number. Then, if both functions are positive definite in their respective domains, the sum $W$ will also be positive definite in the $\R^{n_1+n_2+n_3}$ domain, satisfying the first condition from \Cref{Section:Lyap_stab}. For the second condition, we can take $\dot{W}=\dot{V_1}+\dot{V_2}<0$ as both functions satisfy the condition $\dot{V_i}<0, i \in \{1,2\}$ for appropriate choices of control inputs. 
\end{proof}

\vspace{-12pt}
\hspace*{4mm}Note that this proposition holds when the subsystems S1 and S2 are coupled through the state $x_3$ only and not the controls, i.e., they do not have shared controls.  This will allow taking the corresponding control input to satisfy condition $2$ from \Cref{Section:Lyap_stab} for $V_1$ and $V_2$ individually. However, in most cases, the states are coupled through the controls. To address this case, we proceed as follows:

\vspace{-2pt}
\textbf{Assumption 1.}  Given a dynamical system represented by equation (12) with shared controls and two Lyapunov functions for the subsystems as described in Proposition 1, there exists an admissible control input $u \in \mathcal{U}$ such that the following two conditions are satisfied for all $x=(x_1,x_2,x_3)$: 
\begin{align}
\begin{split}
    \nabla{V_1} \cdot g_1(x_1,x_3,u) &< 0\\
    \nabla{V_2} \cdot g_2(x_2,x_3,u) &< 0 .
\end{split}
\end{align}
Without assumption 1, if the subsystems have shared controls, we could not guarantee that condition 2 from \Cref{Section:Lyap_stab} is satisfied for both $V_1$ and $V_2$. Therefore, $\dot{W}=\dot{V_1}+\dot{V_2}<0$ may not be satisfied for any admissible control. 

\vspace{-6pt}
\hspace*{4mm}Now consider the cost function of RL defined in equation (9) but with our DCLF incorporated: 
\vspace{-6pt}
\begin{equation}\
         V_\gamma^\pi(x_0)= \sum_{k=0}^{\infty} \gamma^k[\Delta W(x_k)+\ell(x_k,\pi(x_k)],
\end{equation}
 where $\Delta W(x_k):=W(f(x_k,\pi(x_k)))-W(x_k)$.
 
 \vspace{-2pt}
 \hspace*{4mm}In the limit of small time steps for the dynamics in equation (1), from Proposition 1, we know that there exists an admissible control that will satisfy $W(f(x_k,\pi(x_k)))-W(x_k)<0$. This means that our incorporated DCLF will motivate the RL algorithm to search for policies that decrease the value of $W$, thereby biasing policy search toward low overall cost and reducing the need to explore all directions of the policy space.

\section{Results}
\label{Section:Results}
\hspace*{4mm}We demonstrate our approach through various examples where a DCLF is computed and then used in an RL algorithm to train a policy for a specific task. As mentioned, this approach works for any RL algorithm since only a modification of the reward function is needed. In our experiments, we demonstrate the method using the state-of-the-art Soft Actor-Critic algorithm (SAC) [32] and the Proximal-Policy Optimization algorithm (PPO) [34]. {Code can be found here: \texttt{https://github.com/CLeARoboticsLab/DCLF-RL}}
\vspace{-6pt}
\subsection{Dubins Car}
\hspace*{4mm}Consider the dynamics of a Dubins Car as follows: 
\begin{align}
\begin{split}
    \dot{x}_1 &= v \cos(x_3) \\
    \dot{x}_2 &= v \sin(x_3) \\
    \dot{x}_3 &= u
\end{split}
\end{align}
where the velocity is a constant at $v=1$, the control is the heading angle given by $x_3$; input bounds are $u \in [-\nicefrac{\pi}{4},\nicefrac{\pi}{4}]$ and the goal is to drive the system to the point ${(x_1,x_2)=(0,0)}$. We restrict the position to the states $x_1,x_2 \in [-5,5]$. 
This system can be decomposed into the following two subsystems: 

\;\;\;\;\;\;\;\;\;\;\;Subsystem S1 \;\;\;\;\;\;\;\;\;\;\;\;\;\;\;\;\;\;\;\;\;Subsystem S2
\begin{align}
  \begin{split}
    \dot{x}_1&= v \cos(x_3) \\
    \dot{x}_3 &= u
  \end{split}
  \;\;\;\;\;\;\;\;\;\;\;
  \begin{split}
    \dot{x}_2 &= v \sin(x_3) \\
    \dot{x}_3 &= u
  \end{split}
\end{align}

\begin{figure}[!t]
\begin{center}
\includegraphics[width=\linewidth]{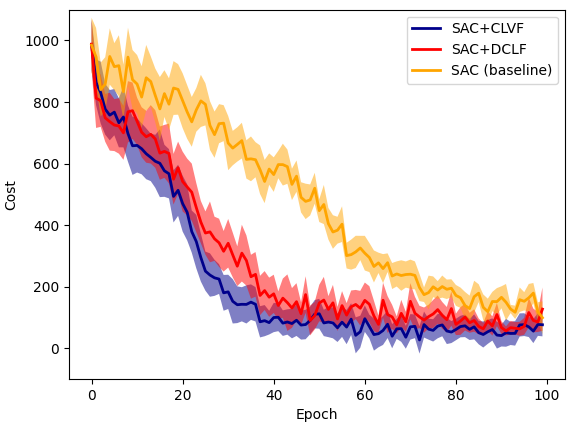}
\end{center}
\caption{Learning curves of different RL algorithms run on a Dubins Car simulation. Approaches incorporating Lyapunov functions, SAC+DCLF (ours) and SAC+CLVF, perform better than the SAC baseline. Each epoch consists of 300 simulation steps or 3 seconds of data. Four different seeds were used in the simulations.}
\label{Fig2}
\end{figure}
\vspace{-6pt}

\hspace*{4mm}Note that the two subsystems are coupled by the third state and the control. As the initial problem has a 3-dimensional state space, it is tractable with HJR, and we can compare our approach with the results of obtaining a CLVF directly on three dimensions. 
We computed a CLVF for each subsystem to obtain the DCLF for the complete system. Next, we incorporated the computed DCLF into the SAC algorithm, where the cost function penalizes the distance to the origin. We benchmark our approach against the standard SAC algorithm and the same RL algorithm incorporating the computed CLVF directly, as shown in Figure 2.  

\vspace{-6pt}
\hspace*{4mm}For the SAC algorithm, a discount factor of $\lambda=0.99$ was used, whereas the RL algorithms with the value function modification allow a smaller discount factor of $\lambda=0.9$. More details about the simulation parameters can be found in the Appendix. With our approach, the algorithm learns a policy in less than $12 \times 10^3$ steps, whereas using standard SAC converges after $3 \times 10^4$ steps. Our approach has a similar performance when compared against the incorporated CLVF, showing no significant difference with the DCLF computed. 

\vspace{-6pt}
\hspace*{4mm}In Fig. 3, we can see a comparison between the trajectories generated by the SAC and SAC+DCLF algorithms after 2 minutes of data, when our approach has converged. As shown, our approach yields a policy that reaches the origin much more quickly than the baseline. 
\begin{figure}[!htbp]
\begin{center}
\includegraphics[width=\linewidth]{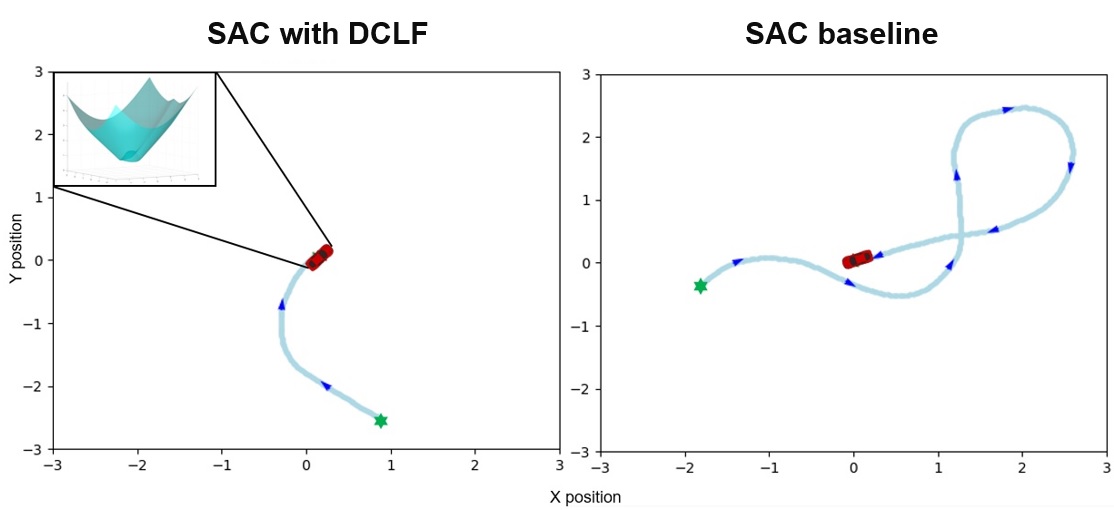}
\end{center}
\caption{Dubins Car after 2 minutes of trajectory data. (Left): Trajectory using our approach with the incorporated DCLF. (Right): Trajectory using standard SAC algorithm.}
\label{Fig3}
\end{figure}

\vspace{-6pt}
\subsection{Lunar Lander}
\hspace*{4mm} In this example, we computed a DCLF for the dynamics of a Lunar Lander from the OpenAI environment [37]. Then, we incorporated it into the SAC and PPO algorithms to learn how to land the vehicle. The system dynamics has six dimensions, which means that a CLF cannot be computed by previous HJR-based methods. The Lunar Lander has state $z=(x,y,v_x,v_y, \theta,\omega)$ and dynamics: 

\begin{equation}
    \begin{bmatrix}
        \dot{x}\\
        \dot{y}\\
        \dot{v_x}\\
        \dot{v_y}\\
        \dot{\theta}\\
        \dot{\omega}
    \end{bmatrix} = \begin{bmatrix}
        v_x\\
        v_y\\
        \frac{1}{m}(-u_1\sin{\theta}+u_2\cos\theta)\\
        \frac{1}{m}(-mg+u_1\cos{\theta}+u_2\sin\theta)\\
        \omega \\
        \frac{\ell_1}{I_c}u_1+\frac{\ell_2}{I_c}u_2+\frac{D}{I_c}\omega
    \end{bmatrix}\,
\end{equation}

\vspace{-4pt}
where $x,y$ represent the horizontal and vertical positions, $\theta$ is the roll angle, $v_x,v_y$ represent the horizontal and vertical velocities, and $\omega$ is the roll rate. The goal can be formally defined as reaching the state $z=(0,0,0,0,0,0)$. The inputs $u_1 \in [0,1]$ and $u_2 \in [-1,1]$ represent the vertical and horizontal thrust applied to the vehicle. The parameters $\ell_1,\ell_2$ are the distance of the center of mass to the respective thrust engine, $D$ is the rotational drag, and $I_c$ is the moment of inertia. This dynamical model is an approximation taken from the equations used in the Box2D environment from [37]. The system can be decomposed into the following two subsystems: 
\vspace{-6pt}
\begin{align}
  \begin{split}
    z_1&=(x,v_x,\theta,\omega)\\
    z_2&=(y,v_y,\theta,\omega).
  \end{split}
\end{align}
\vspace{-2pt}
\hspace*{4mm}We see that the subsystems are coupled by the fifth and sixth dimensions, and they have shared controls. Applying our method, we computed a CLVF for each subsystem to recover the DCLF that is implemented in the cost function from the standard Lunar Lander environment of OpenAI [37]. Fig. 4 shows a comparison between our approach and the standard SAC and PPO algorithms. A discount factor of $\lambda=0.98$ was used for SAC and PPO baselines. With our approach on both algorithms, it was possible to converge to the policy using a discount factor of $0.85$.

\begin{figure}[!h]
\begin{center}
\includegraphics[width=\linewidth]{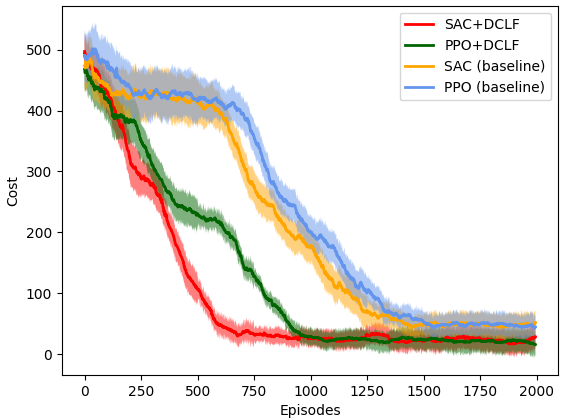}
\end{center}
\caption{Learning curves of different RL algorithms run on the Lunar Lander environment. Five different seeds were used in the simulations.}
\label{Fig4}
\end{figure}

\hspace*{4mm}Our approach with the SAC algorithm achieved the best performance, taking about $700$ episodes to start getting minimum costs, followed by our approach with PPO, which takes $900$ episodes approximately. In contrast, using just SAC and PPO baselines takes about $ 3 \times 10^3$ episodes. 

\vspace{-6pt}
\subsection{Drone}

\hspace*{4mm}In this example, we use a 10-dimensional approximation of a drone's dynamics [39], with state $ {z=(x,y,z,v_x,v_y,v_z,\theta_x,\theta_y,\omega_x,\omega_y)}$ represented by the system in equation \eqref{drone_approx}. The variables $x, y, z$ represent the position, $ v_x, v_y, v_z$ the respective velocities, $ \theta_x, \theta_y$ the pitch and roll of the drone, $ \omega_x, \omega_y$ the pitch and roll rates, and $g$ denotes the acceleration due to gravity. The inputs $u_1 \in [-15,15]$, $u_2 \in [-15,15]$, and $u_3 \in [0,2g]$ represent the desired pitch angle, roll angle, and vertical thrust, respectively. The parameters used are $k_T=1.1$, $p_0=10$, $p_1=8$ and $q_0=10$. 

\begin{equation}
\label{drone_approx}
    \begin{bmatrix}
        \dot{x}\\
        \dot{y}\\
        \dot{z}\\
        \dot{v_x}\\
        \dot{v_y}\\
        \dot{v_z}\\
        \dot{\theta_x}\\
        \dot{\theta_y}\\
        \dot{\omega_x}\\
        \dot{\omega_y}
    \end{bmatrix} = \begin{bmatrix}
        v_x\\
        v_y\\
        v_z \\
        g\tan{\theta_x}\\
        g\tan{\theta_y}\\
        k_Tu_3-g\\
        -p_1\theta_x+\omega_x\\
        -p_1\theta_y+\omega_y\\ 
        -p_0\theta_x+q_0u_1\\
        -p_0\theta_y+q_0u_2\\ 
    \end{bmatrix}\,
\end{equation}

The goal is to land the quadcopter, that is, reach a $0$ value for all the states. The drone is initialized to star at a random position uniformly distributed at least 1.5 meters above the ground but less than 3 meters above, with zero positional and angular velocities.
Here, we can make a decomposition into three subsystems of 4D, 4D, and 2D as follows: 
\begin{align}
  \begin{split}
   z_1&=(x,v_x,\theta_x,\omega_x)\\
   z_2&=(y,v_y,\theta_y,\omega_y)\\
   z_3&=(z,v_z).
    \end{split}
\end{align}
Note that this decomposition does not have shared controls in the subsystems. 

\vspace{-10pt}
\begin{figure}[!h]
\begin{center}
    \includegraphics[width=\linewidth]{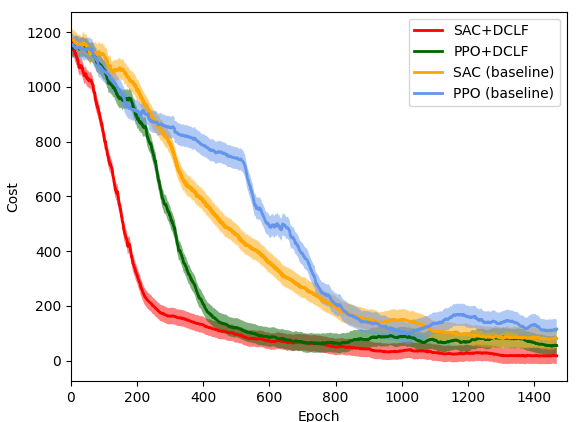}
\end{center}
\caption{Learning curves of different RL algorithms run on the Drone experiment. Each epoch consists of 3 seconds of data. Four different seeds were used in the simulations.}
\label{Fig5}
\end{figure}

\hspace*{4mm}In Fig. 5, we see our simulation results, representing the costs achieved by our approach against the standard SAC and PPO algorithms. The discount factors used were $0.85$ for SAC+DCLF, $0.85$ for PPO+DCLF, $0.97$ for SAC, and $0.95$ for PPO. Our approach with the SAC algorithm achieved the best performance, taking about $1.5 \times 10^3$ seconds to get minimum costs and reach the goal on every simulation episode. Next, our approach with PPO takes approximately the same time to converge to a good policy, although reaching higher costs than our approach with SAC. In contrast, using the SAC baseline takes over $3.6 \times 10^3$ seconds of data to converge, while PPO takes over $4.2 \times 10^3$ seconds. Moreover, they both converge to a policy with higher costs than our approach.

\section{Conclusions and future work}

\hspace*{4mm}By proposing a novel method that exploits system decomposition techniques, we extended previous work in computing Control Lyapunov Functions using Hamilton-Jacobi Reachability analysis. We showed how our computation of the Decomposed Control Lyapunov Function can be incorporated into any standard RL algorithm to alleviate the sample complexity encountered during the training process in RL.\\ 
\hspace*{4mm}We conducted several experiments demonstrating our method by comparing it with state-of-the-art RL algorithms. Our approach allows using a smaller discount factor and finds policies with less data and in less computation time. \\
\hspace*{4mm}A key limitation of our approach is that it presumes knowledge of an accurate dynamics model, which admits a decomposition of the type in Section IV.A. Future work will investigate means of extending these results to other, more general, decompositions, more flexibly coping with shared controls among subsystems, and analyzing the sensitivity of the DCLF to errors in system modeling.

%\printbibliography

\section*{Appendix. Simulation Parameters} 

\hspace*{4mm} We show in Table 1 and Table 2 the simulation parameters used in the RL training process for the different experiments in \Cref{Section:Results}.
\begin{table}[!h]
\centering
    \begin{tabular}{ |p{2.4cm}||p{1.5cm}|p{1.7cm}|p{1.3cm}|  }
     \hline
     \multicolumn{4}{|c|}{SAC algorithm} \\
     \hline
     & Dubins Car & Lunar Lander &Drone\\
     \hline
     Replay Buffer size   & $1 \times 10^4$    & $3 \times 10^4$ & $1 \times 10^5$  \\
     Batch size&   $256$  &  $128$ & $256$ \\
     Optimizer & Adam & Adam & Adam \\
     Actor learning rate    & $3 \times 10^{-4}$ & $3 \times 10^{-4}$ & $3 \times 10^{-4}$ \\
     Critic learning rate& $3 \times 10^{-4}$  & $3 \times 10^{-4}$ & $3 \times 10^{-4}$ \\
     Hidden layers& $2$ &  $3$ & $2$ \\
     Hidden size& $256$  & $128$ & $64$ \\
     \hline
    \end{tabular}
    \caption{Simulation parameters for the different experiments using the SAC algorithm.}
    \label{Table1}
\end{table}
\begin{table}[!h]
\centering
    \begin{tabular}{ |p{2.7cm}||p{2.0cm}|p{1.5cm}|}
     \hline
     \multicolumn{3}{|c|}{PPO algorithm} \\
     \hline
      & Lunar Lander &Drone\\
     \hline
     Replay Buffer size       & $3 \times 10^4$ & $1 \times 10^5$\\
     Batch size  & $128$  & $128$\\
     Policy learn. rate     & $2 \times 10^{-5}$ & $3 \times 10^{-4}$ \\
     Value learn. rate  & $3 \times 10^{-4}$  & $3 \times 10^{-4}$\\
     Hidden layers & $3$ & $2$  \\
     Hidden size  & $64$ & $64$ \\
     \hline
    \end{tabular}
    \caption{Simulation parameters for the different experiments using the PPO algorithm.}
    \label{Table2}
\end{table}

\end{document}